\documentstyle[12pt]{article}
\begin{document}
\title{\Large{{\bf On the zero viscosity limit}}}

\vspace{1.5cm}

\author{~\\{ P. Olesen\footnote{Also at the 
Institute for Advanced Cycling, Blegdamsvej 19, Copenhagen, Denmark}} \\~\\
\it The Niels Bohr Institute\\ \it Blegdamsvej 17\\ \it DK-2100 Copenhagen {\O}
\\ \it Denmark}
\date{\today}
\maketitle
\vfill
\begin{abstract}
The question of whether the zero viscosity limit $\nu\rightarrow 0$
is identical to the no viscosity $\nu\equiv 0$ case is investigated
in a simple shell (GOY) model with only three shells. We find that
it is possible to express two velocities in terms of
Bessel functions. The third velocity function acts as a background.
The relevant Bessel functions are infinitely oscillating 
as $\nu\rightarrow 0$ and are not analytic as functions of $\nu$
at the point $\nu =0$. 
We also mention a perturbative method which may be used to improve the model.

\end{abstract}
\vfill

\newpage

\section{Introduction}

In the Navier-Stokes equation the viscosity $\nu$ plays an important role. 
It is an interesting question of principle whether the limit of 
vanishing viscosity
is given by the solution to the same equation with $\nu$ 
identically zero. In other words, is the limit $\nu\rightarrow 0$
smooth, or is it non-analytic? 
An answer to this question may  be of interest in various
branches of physics, for example in cosmology,
where the history due to the expansion of the Universe may contain different
viscosities and the behavior during transitions from one $\nu$
to another may not be analytic. 

In a numerical study of decaying
turbulence some evidence was found in \cite{axel} that the small
viscosity  limit
is highly non-trivial and does not conform to the naive expectations.
Of course, numerical data do not necessarily allow an extrapolation
simulating the limit $\nu\rightarrow 0$. Therefore it would be
desirable to have some explicit mathematical expression for
the velocity field which allow this limit to be investigated.

It is well known that a direct mathematical study the Navier-Stokes equations
for high Reynolds numbers is not an easy matter. Therefore there is
some motivation for looking for a model of the hydrodynamics equations
where an analytic approach may be more hopeful. To this end one may think of 
the shell (GOY) model of Gledzer, Ohkitani and Yamada\cite{goy}. Many
properties of turbulence, especially those related to energy transfer
and the small intermittency effects, have been understood from the
numerical studies of the shell model. For a review of the applications,
see ref. \cite{mogens}.

The model is formulated in terms of Fourier space velocity variables
$u_n(t)$, and the dynamical equations are given by
\begin{equation}
\left(\frac{d}{dt}+\nu k_n^2\right)u_n^\star=-ik_n\left(u_{n+1}u_{n+2}-
\frac{\delta}{r}~u_{n-1}u_{n+1}-\frac{1-\delta }{r^2}~u_{n-1}u_{n-2}\right)
+f\delta_{n,n_0}.
\label{goy}
\end{equation}
Here $k_n=r^n$, and $f$ is an external force acting on shell number $n_0$
and $n$ is less than some maximum number $N$.
Usually this equation is studied numerically with a large number of shells.
Also, the usual scaling law of Kolmogorov $k_n^{-1/3}$ appears when the maximum
number $N$ of shells go to infinity \cite{mogens}.

There is not much hope that the shell equation (\ref{goy}) can be integrated 
in terms of
standard mathematical functions. Therefore the question of whether the solution
of (\ref{goy}) in the limit $\nu\rightarrow 0$ is identical to the solution
of (\ref{goy}) with $\nu \equiv 0$ would still be subject to numerical
extrapolation.

However, the probability of obtaining explicit solutions for the 
$u_n'$s may increase
if the number of shells is small. In this paper
we shall show that a simple truncated three-shell  model actually
allows a solution in terms of Bessel functions, provided the parameter
$\delta$ takes a somewhat special value. These Bessel functions
turn out to be infinitely oscillating in the limit $\nu\rightarrow 0$,
so that as $\nu =0$ is approached, they can take many different values,
and in general the $\nu\rightarrow 0$ dynamics does not correspond to the
$\nu =0$ case. In some cases it is possible
to ``renormalize'' the bad behavior so as to obtain a smooth limit
for small $\nu$, at the cost of trading a quantity which is
infinite at $\nu=0$ with an initial velocity. 

In section 2 we introduce the three-shell model. As a preliminary we
solve it for $\nu\equiv 0$. Then in section 3 we include viscosity
and solve the model in terms of Bessel functions. In section 4 the
limit $\nu\rightarrow 0$ is investigated. Section 5 contains a discussion of
the large time limit, and section 6 discusses the relatively
simple case where the shell distance is very large. The Bessel functions are 
then approximately replaced by trigonometric functions, which are still 
oscillating infinitely near $\nu =0$. In section 7 we discuss a possible
perturbative improvement of the model, and we conclude in section 8.

\section{The three-shell model}

We shall now consider a simple GOY-model with only three shells, and with 
the forcing term acting on the first shell. Furthermore, we consider the case 
where
\begin{equation}
\delta=1.
\end{equation}
This value means that the dynamics is intermediate between two dimensions,
where $\delta =1+1/r^2$, and three dimensions, where $\delta =1-1/r$. The
choice of $\delta$ is connected to invariants like enstrophy or
helicity invariants, as was first pointed out by Kadanoff et al.
\cite{kadanoff}. See also the review \cite{mogens}. 
The general equations (\ref{goy}) now reduce to the following three equations
\begin{equation}
\left(\frac{d}{dt}+\nu k_1^2\right)u_1^\star(t) =-i k_1u_3(t)u_2(t)+f,
\label{1}
\end{equation}
and
\begin{equation}
\left(\frac{d}{dt}+\nu k_2^2\right)u_2^\star (t)=i k_1u_3(t)u_1(t),
\label{2}
\end{equation}
as well as
\begin{equation}
\left(\frac{d}{dt}+\nu k_3^2\right)u_3^\star (t)=0.
\label{3}
\end{equation}
The last equation simplifies because of $\delta =1$, since otherwise the
right hand side would contain the term $ik_1 (1-\delta) u_1u_2$. 

For future reference we first solve these equations with zero viscosity, 
$\nu =0$. We impose the boundary conditions
\begin{equation}
u_1(0)=0 ~~~{\rm and}~~~u_2(0)\neq 0.
\label{boundary}
\end{equation}
Eq. (\ref{3}) can be trivially solved,
\begin{equation}
u_3 (t)=C_3,
\end{equation}
where $C_3$ is a (complex) constant. Because of this, the velocity field
$u_3(t)$ will act as a background field for the other velocities
$u_1$ and $u_2$. Using this, eq. (\ref{1}), which now reads
\begin{equation}
\frac{du_1^\star (t)}{dt}=-i k_1~C_3~u_2(t)+f,
\label{11}
\end{equation}
can then be integrated,
\begin{equation}
u_1^\star (t)=ft-ik_1~C_3\int_0^tdt'~u_2(t').
\end{equation}
Inserting this solution for $u_1$ in eq. (\ref{2}) with $u_3$ replaced
by $C_3$ we obtain 
\begin{equation}
\frac{d^2u_2^\star (t)}{dt^2}+k_1^2 |C_3|^2~u_2^\star (t)=ik_1C_3f^\star.
\label{diff1}
\end{equation}
This second order differential equation has a simple oscillating solution
which can be inserted in eq. (\ref{11}) in order to obtain $u_1$. Using the
boundary conditions (\ref{boundary}) we simply get
\begin{equation}
u_1^\star (t)=\frac{2iC_3}{|C_3|}~\alpha~\sin k_1|C_3|t ,~~
u_2^\star (t)=2\alpha~\cos k_1|C_3|t~+\frac{iC_3f^\star}{k_1|C_3|^2},
~~{\rm and}~~u_3^\star (t)=C_3^\star.
\label{solution}
\end{equation}
Here $\alpha$ is an integration constant which is related to $u_2$ at $t=0$,
\begin{equation}
u_2^\star(0)=2 \alpha+\frac{iC_3f^\star}{k_1|C_3|^2}.
\end{equation}
It should be noticed that although the forcing was coupled to the equation
(\ref{1}) for $u_1$, in the solutions (\ref{solution}) the constant $f$
occurs also on the second shell.

\section{The three-shell model and viscosity}

We now consider the basic equations (\ref{1})-(\ref{3}) with non-vanishing
viscosity, $\nu \neq 0$. Again, eq. (\ref{3}) can be solved trivially,
\begin{equation}
u_3(t)=C_3~e^{-\nu k_3^2~t}.
\label{background}
\end{equation}
Again this field will act as a background field for the two other velocities.
Inserting eq. (\ref{background}) in (\ref{1}) we can integrate to obtain,
\begin{equation}
u_1^\star (t)=e^{-\nu k_1^2t}~\int_0^tdt'~e^{\nu k_1^2t'}f(t')
-ik_1~C_3 e^{-\nu k_1^2t}\int_0^t dt'~e^{-\nu (k_3^2-k_1^2)t'}~u_2(t').
\end{equation}
Inserting this in eq. (\ref{2}) gives
\begin{eqnarray}
&&\left(\frac{d}{dt}+\nu k_2^2\right)u_2^\star (t)\nonumber \\
&=&-k_1^2|C_3|^2~e^{-\nu (k_1^2+
k_3^2)t}\int_0^tdt'e^{-\nu (k_3^2-k_1^2)t'}u_2^\star (t')+
iC_3 k_1~e^{-\nu(k_1^2+k_3^2)t}~\int_0^tdt'e^{\nu k_1^2t'}f^\star (t') ,
\end{eqnarray}
and by differentiation we obtain the following second order differential
equation,
\begin{equation}
\frac{d^2u_2^\star}{dt^2}+\nu (k_1^2+k_2^2+k_3^2)\frac{du_2^\star}{dt}+
(\nu^2k_2^2(k_1^2+k_3^2)+k_1^2|C_3|^2 e^{-2\nu k_3^2t})u_2^\star=i k_1C_3
f^\star (t) e^{-\nu k_3^2t}.
\label{diff2}
\end{equation}
It should be noticed that formally eq. (\ref{diff2}) reduces to (\ref{diff1})
by taking $\nu =0$. The main point of this paper is that this is
not true for the {\it solution} of (\ref{diff2}), unless time is very small.

In order to solve eq. (\ref{diff2}) it is convenient to introduce
the function $S(t)$,
\begin{equation}
S(t)=e^{\nu (k_1^2+k_2^2+k_3^2)t/2}~u_2^\star (t).
\end{equation}
Eq. (\ref{diff2}) then gives
\begin{eqnarray}
&&\frac{d^2S}{dt^2}+\left(-\frac{\nu^2}{4}(k_1^2+k_2^2+k_3^2)^2
+\nu^2 k_2^2(k_1^2+k_3^2)+k_1^2|C_3|^2e^{-2\nu k_3^2t}\right)S\nonumber \\
&=&ik_1C_3f^\star (t) e^{\frac{\nu}{2} (k_1^2+k_2^2-k_3^2)t}
\label{S}
\end{eqnarray}
The homogeneous equation corresponding to $f=0$ is of the Bessel type, with
the solution
\begin{equation}
S(t)=J_{\pm a}\left(\frac{k_1|C_3|}{\nu k_3^2}e^{-\nu k_3^2t}\right),
\label{bessel}
\end{equation}
where
\begin{equation}
a=\frac{1}{2k_3^2}~\sqrt{(k_1^2+k_2^2+k_3^2)^2-4k_2^2(k_1^2+k_3^2)}=
\frac{1}{2k_3^2}~(k_1^2-k_2^2+k_3^2).
\label{besselindex}
\end{equation}
Since $k_n=r^n$ we see that for large $r$ the index $a$ approaches 1/2.

The inhomogeneous equation (\ref{S}) can be solved by standard methods
from the knowledge of the solution (\ref{bessel}) of the homogeneous
equation,
\begin{equation}
S(t)=\frac{\pi}{2\nu k_3^2\sin \pi a}~ ik_1C_3\left(J_{-a}(z)
\int_c^t dt' J_a(z') g(t')f^\star (t')+J_a(z)\int_t^d dt' J_{-a}(z') 
g(t') f^\star (t')\right),
\label{fullsolution}
\end{equation}
where $c$ and $d$ are arbitrary constants, and where
\begin{equation}
g(t')=e^{-\nu (k_3^2-k_2^2-k_1^2)t'/2}
\label{g}
\end{equation}
and 
\begin{equation}
z=\frac{k_1|C_3|}{\nu k_3^2}e^{-\nu k_3^2t},
\label{z}
\end{equation}
and $z'$ is the same as $z$ except that $t$ is replaced by $t'$.  

It should be noticed that the argument of the Bessel function (\ref{z})
is not nicely behaved as a function of $\nu$. Therefore the formal
coincidence of eqs. (\ref{diff1}) and (\ref{diff2}) for $\nu =0$ is
not reflected in a simple manner in the Bessel solution (\ref{bessel}).

\section{The limit $\nu\rightarrow 0$}

In view of the last remarks we shall now study the limit $\nu\rightarrow 0$. 
To simplify matters, we start by looking at the homogeneous solution
\begin{equation}
u_2^\star (t)=A J_a (z)+B J_{-a} (z),
\label{AB}
\end{equation}
where $A$ and $B$ are integration constants, and where $z$ was defined in eq. 
(\ref{z}). If we maintain the boundary condition $u_1(0)=0$, it follows from
eq. (\ref{2}) that 
\begin{equation}
\left|\frac{du_2^\star}{dt}+\nu k_2^2 u_2^\star\right|_{t=0}=0.
\label{bb}
\end{equation}
From the asymptotic form of the Bessel function valid for $\nu\rightarrow 0$,
i.e. $z\rightarrow\infty$,
\begin{equation}
J_{\pm a} (z)\approx\sqrt{\frac{2}{\pi z}}~\cos (z\mp a\pi/2-\pi/4),
\label{asymptotic}
\end{equation}
and from the explicit expression for $z$ in (\ref{z})
it follows that $u_2^\star$ behaves like $\sqrt{\nu}$ times a cosine.
The time derivative of $u_2^\star$ also behaves like $\sqrt{\nu}$
times a sine, so therefore $\nu k_2^2 u_2^\star$ (which behaves like
$\nu^{3/2}$) is subdominant relative to the time derivative of $u_2$
when $\nu\rightarrow 0$. Therefore we can replace the boundary condition 
(\ref{bb}) by $du_2/dt=0$. This allows us to find the following relation
between the constants $A$ and $B$ in eq. (\ref{AB}),
\begin{equation}
B=-A~\frac{\sin \left(\frac{k_1|C_3|}{\nu k_3^2}-\frac{\pi}{2}a-
\frac{\pi}{4}\right)}{\sin \left(\frac{k_1|C_3|}{\nu k_3^2}+\frac{\pi}{2}a-
\frac{\pi}{4}\right)},
\end{equation}
where we used the asymptotic form (\ref{asymptotic}) for the Bessel functions.
Introducing instead of $A$ the initial value $u_2(0)$, which is assumed to be 
independent of $\nu$, we now get by use of addition theorems for trigonometric
functions
\begin{equation}
A= u_2^\star (0)~\sqrt{\frac{\pi k_1 |C_3|}{\nu k_3^2}}~\frac{\sin\left(
\frac{k_1|C_3|}{\nu k_3^2}+a\frac{\pi}{2}-\frac{\pi}{4}\right)}{\sin (\pi a)}.
\label{renormalization}
\end{equation}
This result allows us to trade the integration constant with the initial 
velocity, and by use of (\ref{asymptotic}) we have
\begin{equation}
u_2^\star (t)\approx u_2^\star (0) e^{-\nu (k_1^2+k_2^2-k_3^2)t/2}~
\cos\left(\frac{k_1|C_3|}{\nu k_3^2}~\left(1-e^{-\nu k_3^2 t}\right)\right).
\label{rr}
\end{equation}
Again we used addition theorems for trigonometric functions.
This result agrees completely with the result (\ref{solution}) for $\nu =0$ 
with $f=0$ if the expansion $\exp (-\nu k_3^2 t) \approx 1-\nu k_3^2 t$ is 
performed. 

In this case it is thus possible to 
absorb the bad behavior for $\nu\rightarrow 0$ in the ``renormalization''
(\ref{renormalization}) of the constant $A$ in terms of $u_2(0)$. The price 
to pay for this
is that the badly behaved constant $A$ (for $\nu\rightarrow 0$ 
$A$ blows up with infinite oscillations) is traded with an initial
velocity assumed to behave nicely as a function of $\nu$.
Therefore, if one does numerical 
simulations it would look as if the limit where $\nu$ is decreased more and 
more does not lead to definite convergent results, unless the
``renormalization'' discussed above is performed at each step of
the limiting procedure.

The basic reason for the violent behavior of $A$ is that the Bessel
function (\ref{bessel}) is not analytic as a function of $\nu$ in 
the point $\nu =0$. Here the most singular behavior is
\begin{equation}
\frac{\partial J_{\pm a}(z)}{\partial\nu}\propto -\frac{k_1 |C_3|}
{\nu^{3/2}k_3^2}
e^{-\nu k_3^2 t}~\cos\left(\frac{k_1|C_3|}{\nu k_3^2}e^{-\nu k_3^2 t}
\mp a\frac{\pi}{2}-\frac{\pi}{4}\right)+O\left(\frac{1}{\sqrt {\nu}}\right).
\end{equation}
This shows that the solution of the differential equation (\ref{diff2})
does not behave in a simple way for small $\nu$.

It is not always possible to ``renormalize'' the bad
behavior away. The function
\begin{equation}
u_2^\star (t)=u_2^\star (0)e^{-\nu (k_1^2+k_2^2+k_3^2)t/2}J_a (k_1|C_3|/
(\nu k_3^2)~e^{-\nu k_3^2 t})/J_a (k_1|C_3|/(\nu k_3^2))
\end{equation}
is a solution which rapidly oscillating for $\nu\rightarrow 0$, and no
``renormalization'' trick can remove this behavior. 
We thus see that the simple three-shell model can produce a variety
of results which look rather non-trivial.

Instead of imposing some initial velocity for the $u_1$ and $u_2$ fields,
we can proceed in a more physical way by assuming that initially 
$u_1=u_2=0$ up to the time $t_0$.  Then some external agency applies
a force for a limited time from $t_0$ to $t_1$, so after the time $t_1$ the 
solution of the homogeneous equations of motion emerges, as can be seen from
eq. (\ref{fullsolution}) with $f(t)$ inside the integrals. However, if
we write this solution as
\begin{equation}
S(t)=A~J_{a} (z)+B~J_{-a} (z)~~{\rm for}~~t>t_1,
\end{equation}
then $A$ and $B$ are not arbitrary coefficients to be fixed by
some initial velocity of $u_1$ and/or $u_2$. On the contrary, these constants 
are fixed dynamically by the force and by the initial velocity $C_3$
of $u_3$,
\begin{equation}
A=\frac{-i\pi k_1C_3}{2~\nu ~k_3^2~ \sin \pi a}~\int_{t_0}^
{t_1}dt J_{-a} (z)~g(t)~f^\star (t)
B=\frac{i\pi k_1C_3}{2~\nu ~k_3^2~ \sin \pi a}~\int_{t_0}^{t_1}dt J_a (z)~g(t)~
f^\star (t).
\label{forceinduced}
\end{equation}
Here there is no way of ``renormalizing'' the constants. If the force
acts in a short time, the situation is similar to what was discussed
above. For example, if we take a delta-function pulse,
\begin{equation}
f(t)=f_0~\delta (t-t_\star),
\end{equation}
we obtain for $\nu$ so small that $g(t)\approx 1$
\begin{equation}
S(t)\approx \frac{-iC_3f_0^\star}{|C_3|}~\sin \left(\frac{k_1|C_3|}{\nu k_3^2}
\left(e^{-\nu k_3^2 t}-e^{-\nu k_3^2 t_\star}\right)\right)\theta(t-t_\star ).
\label{tt}
\end{equation}
In arriving at this result we used standard addition formulas for
the trigonometric functions.
For $\nu$ small and time not too large this again gives the $\nu =0$
result, which in the present case is arrived at smoothly.

The situation changes completely if we let the force act for a longer time. 
Then the integrals in eq. (\ref{forceinduced}) in general involve
times where an expansion like $e^{-\nu k_3^2 t}\approx 1-\nu k_3^2 t$
is not valid {\it inside} the integrals. Therefore a result analogous to 
(\ref{tt}) with a smooth $\nu\rightarrow 0$ limit does not appear if
the force is allowed to act long enough. As a matter of fact, the integrals
determining $A$ and $B$ in (\ref{forceinduced}) are strongly oscillating, as 
one can see in numerical examples.

\section{The large time limit}

In the case where $t\rightarrow \infty$ in such a way that $1/\nu~e^
{-\nu k_3^2 t}\rightarrow 0$ only the lowest order terms in the Bessel 
functions should be kept, $J_a\propto z^a (1+O(z^2))$, so for $f=0$ we have
\begin{equation}
u_2^\star (t)\approx L_-~e^{-\nu b_- t/2}\left(1-\frac{k_1^2|C_3|^2}
{4\Gamma (2-a)}~e^{-2\nu k_3^2t}\right)+L_+~e^{-\nu b_+ t/2}\left(1-
\frac{k_1^2|C_3|^2}{4\Gamma (2+a)}~e^{-2\nu k_3^2t}\right),
\label{u2}
\end{equation}
where the $L'$s are constants and
\begin{equation}
b_{\pm}=k_1^2+k_2^2+k_3^2\pm~2k_3^2a,
\end{equation}
which means
\begin{equation}
b_+/2=k_1^2+k_3^2~~{\rm and}~~b_-/2=k_2^2.
\label{bs}
\end{equation}

The function $u_1(t)$ can be obtained most simply directly from eq. 
(\ref{2}). The result is again expressed in terms of Bessel functions.
We can  find the asymptotic behavior corresponding to eq. (\ref{u2})
by inserting eq. (\ref{u2}) in eq. (\ref{2}). From (\ref{bs}) we
have  for $t$ large
\begin{equation}
u_2^\star (t)\approx L_-~e^{-\nu k_2^2 t}\left(1-\frac{k_1^2|C_3|^2}
{4\Gamma (2-a)}~e^{-2\nu k_3^2t}\right)+L_+~e^{-\nu (k_1^2+k_3^2) t}\left(1-
\frac{k_1^2|C_3|^2}{4\Gamma (2+a)}~e^{-2\nu k_3^2t}\right).
\label{u22}
\end{equation}
The leading term contains $e^{-\nu k_2^2 t}$ as one would expect.
It is important that this term is
annihilated by the operator $d/dt+\nu k_2^2$ in eq. (\ref{2}). By use of
eq. (\ref{2}) the result for $t$ large is thus
\begin{eqnarray}
u_1(t)&\approx& \frac{1}{ik_1C_3}(-\nu L_+(k_1^2-k_2^2+k_3^2)~e^
{-\nu k_1^2t} +\frac{\nu L_-k_3^2k_1^2 |C_3|^2}{2\Gamma (2-a)}~
e^{-\nu (k_2^2+k_3^2)t}\nonumber \\
&+&\nu\frac{L_+k_1^2|C_3|^2(3k_3^2-k_2^2+k_1^2)}{4\Gamma
(2+a)}~e^{-\nu (k_1^2+2k_3^2)t})
\end{eqnarray}
As one would expect, the leading term is simply $e^{-\nu k_1^2t}$,
which is annihilated by the operator $d/dt+\nu k_1^2$.
This $u_1$ can be reinserted in eq. (\ref{1}) as a check of the 
self-consistency of eqs. (\ref{1}) and (\ref{2}).

\section{Simplified results for $r\gg 1$}

We shall now mention that in the limit where the shell distance $r$ is
very large, our results simplify considerably. From eq. (\ref{besselindex})
we obtain $a\approx 1/2$, since $k_1^2$ and $k_2^2$ can be ignored relative
to $k_3^2$, so the solution can now be expressed in terms of
trigonometric functions,
\begin{eqnarray}
u_2^\star(t)&=&k_3\sqrt{\frac{\nu}{\pi k_1 |C_3|}}~
e^{-\nu (k_1^2+k_2^2)t/2}
\left[A~\sin\left(
\frac{k_1|C_3|}{\nu k_3^2}e^{-\nu k_3^2t}\right)+B~\cos\left(
\frac{k_1|C_3|}{\nu k_3^2}e^{-\nu k_3^2t}\right)\right]\nonumber \\
&+&{\rm inhomogeneous}~f~{\rm term}.
\label{100}
\end{eqnarray}
From eq. (\ref{2}) we can then find the corresponding field $u_1(t)$,
\begin{eqnarray}
u_1(t)&=&\frac{ik_3}{k_1C_3}~\sqrt{\frac{\nu}{\pi k_1 |C_3|}}~
e^{\nu (2k_3^2-k_1^2-k_2^2)t/2}\nonumber \\
&&
\times[(-A\frac{\nu}{2}(k_1^2-k_2^2)+k_1|C_3|B~e^{-\nu k_3^2t})
\sin\left(\frac{k_1|C_3|}{\nu k_3^2}e^{-\nu k_3^2t}\right) \nonumber \\
&&+(-B\frac{\nu}{2}(k_1^2-k_2^2)-Ak_1|C_3|e^{-\nu k_3^2t}
+\nu k_1^2B)\cos\left(\frac{k_1|C_3|}
{\nu k_3^2}e^{-\nu k_3^2t}\right)]\nonumber \\
 &+&{\rm inhomogeneous}~f~{\rm term}.
\label{101}
\end{eqnarray}
Again we can fix the integration constants by suitable boundary conditions.
Also, we see that the limit $\nu\rightarrow 0$ is not well defined due
to the non-analytic behavor of the arguments of the sine and cosine,
which oscillate violently as $\nu$ is decreased.

\section{Perturbations in $1-\delta$}

To improve the approach presented above one could try a perturbative
expansion in $(1-\delta )$. Denoting the $\delta =1$ functions found above by
$u_1^{(0)}$ and $u_2^{(0)}$ eq. (\ref{3}) would change into
\begin{equation}
\left(\frac{d}{dt}+\nu k_3^2\right)u_3^\star (t)=ik_1 (1-\delta)~ u_1^{(0)}
u_2^{(0)}.
\label{pert}
\end{equation}
This is an equation for $u_3$ with a time-dependent ``forcing'' term. It
can be solved for $u_3$ in terms of the unperturbed functions $u_1^{(0)}$
and $u_2^{(0)}$. This perturbed $u_3$ should then be inserted in eqs.
(\ref{1}) and (\ref{2}) to give the perturbed $u_1$ and $u_2$. Of course,
there is no guarantee that such an expansion is convergent.

From eq. (\ref{pert}) we easily obtain
\begin{equation}
|u_3|^2\approx e^{-2\nu k_3^2t}~\left[|C_3|^2+ik_1(1-\delta )
\int_0^t dt'e^{\nu k_3^2t'}~\left(C_3u_1^{(0)}(t')u_2^{(0)}(t')-C_3^\star
(u_1^{(0)}(t'))^\star (u_2^{(0)}(t'))^\star\right)\right].
\label{ptt}
\end{equation}
If we use eq. (\ref{1}) to express $u_2$ in terms of $u_1$, so we have
\begin{equation}
C_3u_1^{(0)}(t)u_2^{(0)}(t)-C_3^\star (u_1^{(0)}(t))^\star 
(u_2^{(0)}(t))^\star = \frac{ie^{\nu k_3^2t}}{k_1}\left[\frac{d|u_1^{(0)}|^2}
{dt}+2\nu k_1^2|u_1^{(0)}|^2-fu_1^{(0)}-f^\star u_1^{(0)\star}\right].
\end{equation}
This expression can be inserted ie eq. (\ref{ptt}) and after a partial 
integration we obtain
\begin{equation}
|u_3|^2\approx e^{-2\nu k_3^2t}~\left[|C_3|^2-(1-\delta )\left(e^{2\nu k_3^2t}
|u_1^{(0)}(t)|^2-|u_1^{(0)}(0)|^2\right)+(1-\delta ) I(t)+
(1-\delta )J(t)\right],
\label{pertc}
\end{equation}
where we defined
\begin{equation}
I(t)=2\nu (k_3^2-k_1^2)~\int_0^tdt'~e^{2\nu k_3^2t'}~|u_1^{(0)}(t')|^2,
\label{I}
\end{equation}
and
\begin{equation}
J(t)=\int_0^tdt'e^{2\nu k_3^2t'}~\left(f u_1^{(0)}(t')+f^\star (u_1^{(0)}
(t'))^\star \right).
\label{J}
\end{equation}
It is interesting that the third term in the square bracket in eq.
(\ref{pertc}) containing the integral $I$ has a definite sign depending
on $1-\delta $: If $1-\delta >0$ the sign is positive since
$k_3>k_2$. This $\delta$ corresponds to helicity ($=\sum (-1)^n k_n|u_n|^2$)
conservation for $\nu =f=0$. So
through the $I-$term the two first shells give a positive contribution to
the energy of the third shell in the lowest order perturbation theory, 
which means
transfer of energy from lower to higher $k'$s. This is precisely what would
be expected in three dimensions. 

On the other hand, if $1-\delta <0$, the $I-$contribution is negative, and 
the energy in the third shell is decreased by this effect, as expected in two
dimensions where the enstrophy ($=\sum k_n^2|u_n|^2$) is conserved 
for $\nu =f=0$.

Of course, in the full expression (\ref{pertc}) there are two other terms
proportional to $1-\delta$. For $f=0$ it should be possible
 to investigate all the terms
in (\ref{pertc}) numerically by inserting a Bessel function constructed from
the equation of motion (\ref{2}) and the solution for $u_2$, thereby giving
$u_1$. It would be interesting to see if  the overall
sign is $+(1-\delta )$ corresponding to the expected inverse cascade 
(transfer of energy from shorter to larger scales, i.e. from larger to 
smaller $k'$s) in ``two dimensional'' systems ($\delta =1+1/r^2 >1$ ) and 
a forward cascade in ``three dimensions'' ($\delta =1-1/r <1$) where the 
energy is transported from smaller to larger $k'$s. 

It should be noted that if $r$ is large, $\delta$ is in both cases
close to 1. Consequently, if the three-shell model should attempt
to be somewhat similar to the GOY model with many
shells, this will  work best for large separations between the shells.

\section{Conclusions}

In the simple three-shell model we have found the velocity functions
in terms of Bessel functions with an argument which is not
analytic in the viscosity $\nu$. In some cases it is possible to hide this
singular behavior by a suitable ``renormalization''. However, this is not 
true in all cases. So the model may never approach the
similar model with no viscosity, $\nu\equiv 0$, even if $\nu\rightarrow 0$

Of course, the reason for the integrability of the three-shell model is
that the third shell's velocity becomes a fixed background field for 
$u_1$ and $u_2$. Thereby the complexity due to the basic non-linearity
of the shell model has disappeared or, at best, become rudimentary. 
However, this does not make the model
completely trivial, since there is a non-trivial coupling between
$u_1$ and $u_2$. So different Fourier modes do couple, and the coupling
makes transfer of energy between these modes possible. 

We also discussed a perturbation approach with an expansion in
$1-\delta$ around $\delta =1$ where the three-shell model was
originally defined. Although the resulting perturbative change of
the energy of the third mode is relatively complicated, we identified one term
which has the expected cascade properties between three and four
dimensions. We hope to be able to perform an precise analysis of the
sign of all the terms later.

In conclusion one can say that although the three-shell model is
of course an immense simplification of the GOY model, nevertheless
it has features which indicate a non-trivial dynamics.

\vspace{.2cm}

{\bf Acknowledgement:}

I thank Mogens H\o gh Jensen for many interesting  discussions over several 
years.

\end{document}